\documentclass[journal,10pt]{IEEEtran}
\usepackage{geometry}
\usepackage{graphicx}
\usepackage[fleqn]{amsmath}
\usepackage{cases}
\usepackage{subfigure}
\usepackage{stfloats}
\usepackage{epstopdf}
\usepackage{cite}
\usepackage{ulem}
\usepackage{algorithm}
\usepackage{algorithmic}
\usepackage{tabu}

\usepackage{amssymb}
\usepackage{color}
\usepackage{xcolor}
\begin{document}
\title{Enhanced Sequential Covariance Intersection Fusion}

\author{Zhongyao~Hu, Bo~Chen, Wenan~Zhang, Li~Yu% <--this % stops a space
%\thanks{Manuscript received XXX; accepted XXX. Date of publication XXX; date of current version XXX. This work was supported in part by the National Natural Science Funds of China under Grant 61973277 and Grant 61673351, and in part by the Zhejiang Provincial Natural Science Foundation of China under Grant LR20F030004.}
% <-this % stops a space
\thanks{Z. Hu, B. Chen, W. Zhang and L. Yu are with Department of Automation, Zhejiang University of Technology, Hangzhou 310023, China. (email: bchen@aliyun.com).}}% <-this % stops a space

\markboth{}
{Shell \MakeLowercase{\textit{et al.}}: Bare Demo of IEEEtran.cls for Journals}

\maketitle

\begin{abstract}
This paper is concerned with the sequential covariance intersection (CI) fusion problem that the fusion result is independent of fusion structure including the fusion order and the number of estimates fused in each sequential fusion. An enhanced sequential CI fusion is first developed to better meet the practical requirements as compared with the existing batch and sequential CI fusion. Meanwhile, it is proved that the enhanced sequential CI fusion ensures the fusion estimate and covariance are unbiased and consistent. Notice that the fusion structure of the enhanced sequential CI fusion is unpredictable in practice, which may have negative impacts on the fusion performance. To this end, a weighting fusion criterion with analytical form is further proposed, and can be depicted by different formulas when choosing different performance indexes. For this criterion, it is proved that the fusion results are not affected by the fusion structure, and thus the fusion performance can be guaranteed. Finally, simulation examples are utilized to demonstrate the effectiveness and advantages of the proposed methods.
\end{abstract}


\vspace{-3pt}
\section{Introduction}
In distributed fusion estimation, the sensors first filter their own raw measurements to produce local estimates, and then transmit them to the fusion node (FN) for subsequent processing. Due to its flexible structure and good robustness, distributed fusion estimation has received a great deal of attention in the fields of localization \cite{9057730}, medicine \cite{DFKLF} and lithium-ion battery \cite{LIN2017560}. Furthermore, distributed fusion can be divided into two types depending on the data fusion process: batch fusion and sequential fusion. Under the batch fusion as shown in Fig. 1(a) \cite{ZHANG2018358}, the FN fuses the local estimates together after all of them have been received. By doing so, the computational burden is concentrated at the moment when the last local estimate is received, and thus computational delay may be induced. In contrast, under the sequential fusion as shown in Fig. 1(b), the FN receives and fuses the local estimates in parallel. In this case, the computational resources are utilized more efficiently and the computational delay can be reduced. Particularly, the optimal batch and sequential fusion algorithms in the sense of minimum mean square error (MMSE) have been presented in \cite{SUN20041017,ZHANG2018358} respectively, and it was shown in \cite{ZHANG2018358} that the sequential and batch MMSE fusion algorithms have the same accuracy. However, implementing both the batch and sequential MMSE fusion algorithms requires knowledge of the correlations (i.e., cross-covariances) among local estimates, which are difficult to be obtained in practice. Therefore, it is more practical to study distributed fusion estimation with unknown correlations.
    \begin{figure}[thpb]
      \centering
      \includegraphics[scale=0.27]{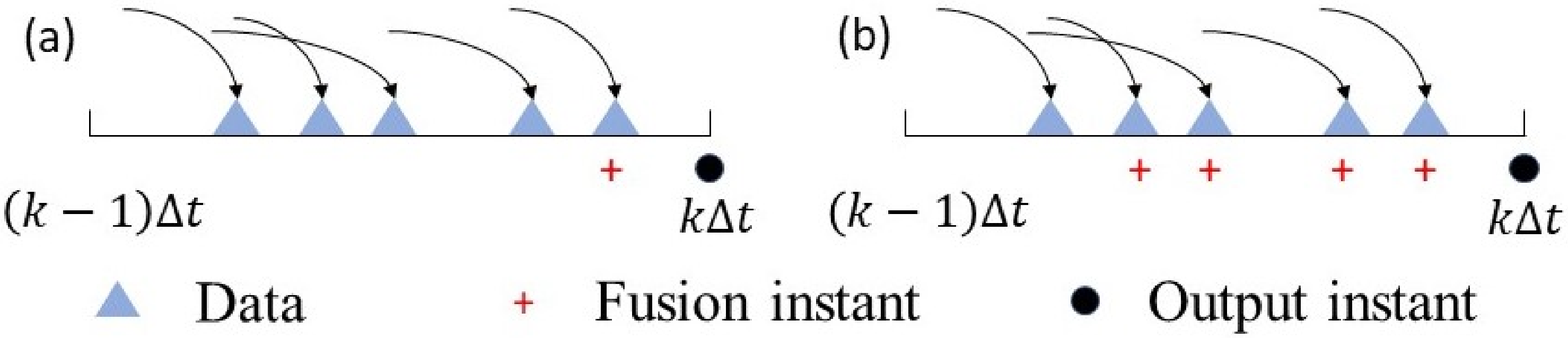}
      \caption{(a). The time shaft corresponding to the batch fusion; (b). The time shaft corresponding to the sequential fusion discussed in \cite{LIN2019128,YAN20133607,Zhang8776655}.}
      \label{figurelabe1}
    \end{figure}

At present, the most popular method is covariance intersection (CI) fusion when the correlations are unknown. The key advantage of CI fusion is that, it can provide a conservative covariance which overestimates the true covariance of the fusion estimate without knowing correlations \cite{Reinhardt7005422}. However, most of the existing works focus on the batch CI (BCI) fusion \cite{Niehsen1020907,Chen1047015,Marc7005422,Julier609105,Franken1591849,Hurley1021196,Wang6129634,BATTISTELLI2014707,Duan6916177,Tang8519318}, while only few works aim to studying the sequential CI (SCI) fusion \cite{DENG2012293,Wang7531499,Wang8028191,CONG201628}. By embedding the classical BCI fusion algorithm \cite{Julier609105} into the sequential fusion framework, a series SCI fusion algorithms were proposed in \cite{DENG2012293,Wang7531499,Wang8028191}. Notice that, due to the differences in deployment and computation speed of different sensor nodes, the moments that the FN receives the local estimates are unpredictable. Under this case, the fusion accuracy of the algorithms in \cite{DENG2012293,Wang7531499,Wang8028191} cannot be guaranteed and may deteriorate. Moreover, in the popular consensus scheme, all nodes are required to have the same state \cite{BATTISTELLI2014707}. This is where it becomes very necessary to eliminate the effect of fusion order on fusion results. For this reason, a sequential fast CI (SFCI) fusion algorithm that is insensitive to the fusion order was proposed in \cite{CONG201628} by modifying the weights. However, to achieve this purpose, it assigns larger weights to those local estimates with little reliability, which is clearly not reasonable. Furthermore, when utilizing the SCI fusion algorithms in \cite{DENG2012293,Wang7531499,Wang8028191,CONG201628}, the FN has to perform a fusion for each local estimate received. In fact, during the actual communication, the FN may receive multiple packets in a flash, in which case the above SCI fusion algorithms may no longer meet the requirements.

Motivated by the above analysis, this paper aims to investigate the CI fusion under sequential fusion framework such that fusion results are not affected by the fusion structure including the fusion order and the number of estimates fused in each sequential fusion. The main contributions of this paper are summarized as follows:
\begin{itemize}
\item An enhanced SCI (ESCI) fusion is developed to be more universal than the existing BCI and SCI fusion, and it is proved that the proposed ESCI fusion has the unbiasedness and consistency.
\item A weighting fusion criterion with analytical form is designed such that the fusion result is independent of the fusion structure. Moreover, this criterion satisfies the principle of importance allocation and allows different importance indicators to be selected for satisfying different requirements.
\end{itemize}
Finally, three simulation examples are used to show the advantages of the proposed methods.

\textbf{Notations:} ${\mathbb{R}}^r$ and ${\mathbb{R}}^{r\times s}$ denote the $r$-dimensional and $r\times s$ dimensional Euclidean spaces, respectively. ${E}(\cdot)$ denotes mathematical expectation. $\mathrm{diag}(\cdot)$ stands for block diagonal matrix. $O$ is zero matrix. $I$ stands for identity matrix. $\mathrm{Tr}(\cdot)$ and $\mathrm{Det}(\cdot)$ represent the trace and determinant of matrix, respectively. $A>B$ and $A\geq B$ imply that $A-B$ is a semi-positive definite and positive definite matrix respectively.
\section{Problem Formulation}
Consider the scenario that a quantity $x\in\mathbb{R}^{d_x}$ is estimated by multiple sensor nodes. Each sensor node can produce a local estimate and a local covariance. Suppose $n$ available local estimate pairs (EPs) $\{x_j,P_j\}$, $j=1,2,\cdots,n$ are to be fused at a FN, where $x_j\in\mathbb{R}^{d_x}$ and $P_j\in\mathbb{R}^{d_x\times d_x}$ denote the $j$th local estimate and covariance, respectively.

The data fusion process of BCI fusion is plotted in Fig. 2(a). In this case, the FN fuses local EPs together after receiving all of them. Particularly, the mathematical expression for the BCI fusion is \cite{Julier609105}
\begin{equation}\begin{aligned}
\left\{ \begin{array}{l}
x^b=\sum^{n}_{j=1}w^{b}_{j}P^bP^{-1}_{j}x_{j}\\ P^b=(\sum^{n}_{j=1}w^{b}_{j}(P_{j})^{-1})^{-1},
\end{array} \right.
\label{eq:14}
\end{aligned}\end{equation}
where $\{x^b,P^b\}$ is the fusion EP of BCI fusion and the weights satisfy $\sum^n_{j=1}w^b_j=1$, $0\leq w^b_j\leq 1$. However, The BCI fusion tends to waste computational resources and induce computational delay, which needs to be avoided.

Let the FN receive the local EPs in the order $\{x_{r_1},P_{r_1}\}$, $\{x_{r_2},P_{r_2}\}$, $\cdots$, $\{x_{r_n},P_{r_n}\}$, and then the data fusion process of SCI fusion is shown in Fig. 2(b). As can be seen from Fig. 2(b), the FN performs a fusion for each local EP it receives. In this case, the computational resources are utilized more efficiently and computational delay can be reduced. Moreover, the SCI fusion can be formulated as \cite{DENG2012293}
\begin{equation}\begin{aligned}
&\left\{ \begin{array}{l}
x^s_{i}=w^{s}_{i}P^s_{i}(P^s_{i-1})^{-1}x^s_{i-1}\\
\ \ \ \ \ \ \ \ +w^s_{i,i}P^s_{i}P^{-1}_{r_i}x_{r_i}\ \ \ \ \ \ \ \ \ \ \ \ \ \ \ \ \ \ i=1,\cdots,n\\
P^s_{i}=(w^{s}_{i}(P^s_{i-1})^{-1}+w^{s}_{i,i}P^{-1}_{r_i})^{-1}
\end{array} \right.
\end{aligned}\end{equation}
where $\{x^s_i,P^s_i\}$ is the fusion EP when the $i$th SCI fusion is performed. The weights satisfy $w^s_i+w^s_{i,i}=1$, $0\leq w^s_i\leq1$ and $0\leq w^s_{i,i}\leq 1$. Meanwhile, $w^s_1\triangleq0$, $x^s_0\triangleq O$ and $(P^s_0)^{-1}\triangleq O$.
%In this case, the computational resources are utilized more efficiently and computational delay can be reduced.

From the above analysis we can see that the BCI fusion (1) and the SCI fusion (2) are extreme. Specifically, (1) fuses all $n$ local EPs at once and (2) allows only one local EP to be fused at a time. In fact, it is possible for the FN to receive multiple local EPs in a flash. Under this case, the BCI and SCI fusion is no longer applicable.
%For example, when the FN receives multiple local EPs in a flash, it is better to fuse these local EPs at once.
%, for example, the FN receives multiple local EPs in a flash.
Therefore, an enhanced SCI (ESCI) fusion is developed in this paper, and its data fusion process is shown in Fig 2(c).
%The data fusion process of ESCI fusion is shown in Fig. 2(c), from which we can be known that, under the ESCI fusion framework, the FN fuses local EPs in the form of ``multiple by multiple" rather than ``one by one" as before.
%However, multiple packages may be transmitted to the FN at a moment. In this case, a ESCI fusion is developed in this paper to better meet the actual requirements. Let the FN receive the local EPs in the order $\{x_{r_1},P_{r_1}\}$, $\{x_{r_2},P_{r_2}\}$, $\cdots$, $\{x_{r_n},P_{r_n}\}$, where $\{r_1,r_2,\cdots,r_n\}=\{1,2,\cdots,n\}$. Then, the data fusion process of the SCI fusion is drawn in Fig. 2(b).
%From this figure we can see that the ESCI fusion allows the local EPs fused in the form ``one batch after another".
Suppose ESCI fusion are performed $t$ times to fuse the $n$ local EPs, and then the mathematical expression for the ESCI fusion can be written as
\begin{equation}\begin{aligned}
&\left\{ \begin{array}{l}
x^e_{i}=w^{e}_{i}P^e_{i}(P^e_{i-1})^{-1}x^e_{i-1}\\
\ \ \ \ \ \ \ \ +\sum^{b_i}_{j=b_{i-1}+1}w^{e}_{i,j}P^e_{i}P^{-1}_{r_j}x_{r_j}\\
P^e_{i}=(w^{e}_{i}(P^e_{i-1})^{-1}\\
\ \ \ \ \ \ \ \ +\sum^{b_i}_{j=b_{i-1}+1}w^{e}_{i,j}P^{-1}_{r_j})^{-1}
\end{array} \right.\ \ i=1,\cdots,t
\end{aligned}\end{equation}
where $\{x^e_i,P^e_i\}$ is the fusion EP when $i$th ESCI fusion is performed. $b_i\triangleq\sum^i_{j=1}a_j$, where $a_i$ is the number of local EPs fused in the $i$th ESCI fusion. Moreover, $b_0\triangleq0$, $w^e_1\triangleq0$, $x_{e,0}\triangleq O$, $P^{-1}_{e,0}\triangleq O$ and the weights $w^e_i$ and $w^e_{i,j}$ should satisfy
\begin{equation}\begin{aligned}
\left\{ \begin{array}{l}
0\leq w^e_i\leq 1,\ 0\leq w^e_{i,j}\leq 1\\
w^{e}_{i}+\sum^{b_i}_{j=b_{i-1}+1}w^{e}_{i,j}=1.
\end{array} \right.
\end{aligned}\end{equation}
    \begin{figure}[thpb]
      \centering
      \includegraphics[scale=0.26]{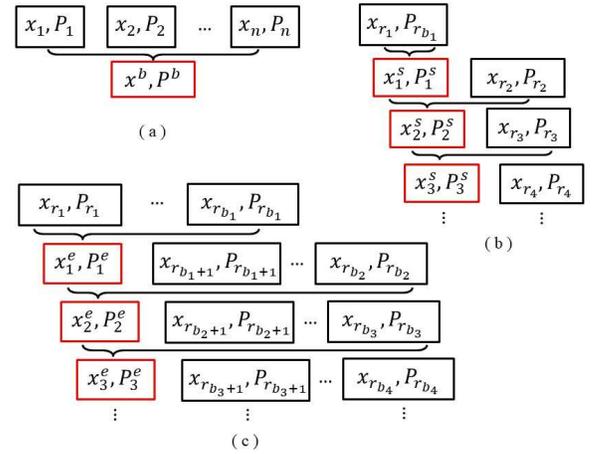}
      \caption{(a). Then data fusion process of the BCI fusion; (b). The data fusion process of the SCI fusion; (c). The data fusion process of the ESCI fusion.}
      \label{figurelabe2}
    \end{figure}

Before proceeding further, define the series $r\triangleq\{r_1,r_2,\cdots,r_n\}$ and $a\triangleq\{a_1,a_2,\cdots,a_t\}$. Here, $r$ is used to represent the order that the FN receives the local EPs, $a$ is used to denote the number of local EPs fused in each ESCI fusion, and $r$ and $a$ are referred to as fusion structure. Obviously, when $a=\{n\}$ and $a=\{1,1,\cdots,1\}$, the proposed ESCI fusion framework (3) will degenerate to the BCI fusion (1) and SCI fusion (2), respectively. This fact means that the ESCI fusion is more universal than the BCI and SCI fusion, and thus it can better meet the practical requirements. Moreover, notice that the unbiasedness and consistency are the most important properties possessed by the BCI and SCI fusion. In this case, it is necessary to prove the developed ESCI fusion also has the unbiasedness and consistency.

On the other hand, the moments of FN receives the local EPs are unpredictable in practice. As a result, the fusion order of local EPs (i.e., $r$) and the number of local EPs fused in each ESCI fusion (i.e., $a$) are unpredictable as well. Then, if $\{x^e_t,P^e_t\}$ varies with $r$ and $a$, the fusion performance is necessarily not guaranteed and even deteriorates. Therefore, it is desired that $\{x^e_t,P^e_t\}$ is not affected by $r$ and $a$.

Based on the above analysis, the problems to be solved in this paper are summarized as:
\begin{itemize}
\item Prove the developed ESCI fusion (3) has the unbiasedness and consistency.
\item Design $w^e_i$ and $w^e_{i,j}$ such that $\{x^e_t,P^e_t\}$ is independent of $r$ and $a$.
\end{itemize}

\textbf{Remark 1:} An example is provided to explain the fusion structure. Assume 4 local EPs are available in a FN. When $r=\{1,2,3,4\}$, $a=\{3,1\}$, the fusion structure is shown in Fig. 3(a) and the fusion EP is denoted as $\{x(1),P(1)\}$. When $r=\{4,2,1,3\}$ and $a=\{2, 2\}$, the fusion structure is shown in Fig. 3(b) and the fusion EP is represented as $\{x(2),P(2)\}$. Moreover, when $r$ and $a$ take other values, there will also be $\{x(i),P(i)\}$, $i=3,4,\cdots$. Since the fusion structure is unpredictable, we cannot determine in advance which fusion EP of the FN will be. In this case, it is desired that the final fusion EP should be independent of $r$ and $a$, i.e., let $x(i)=x(j)$ and $P(i)=P(j)$ $(\forall i,j \in \{ 1,2, \cdots \})$ for different $r$ and $a$.
    \begin{figure}[thpb]
      \centering
      \includegraphics[scale=0.27]{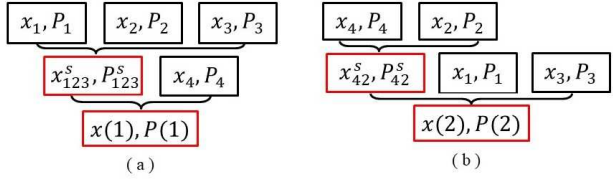}
      \caption{(a). The fusion structure corresponding to $r=\{1,2,3,4\}$, $a=\{3,1\}$; (b). The fusion structure corresponding to $r=\{4,2,1,3\}$ and $a=\{2,2\}$;}
      \label{figurelabe3}
    \end{figure}
\section{Main Results}
%As pointed out in the previous discussion, BCI fusion framework (1) is widely accepted because it provides a consistent fusion covariance. Moreover, unbiasedness is also an good property possessed by (1). Therefore, it will be shown later that the ESCI fusion framework (2) also guarantees consistent and unbiased fusion EP.

\textbf{Theorem 1:} The ESCI fusion (3) can provide the unbiased and consistent fusion EP, i.e., $\{x^e_t,P^e_t\}$ satisfies the following properties:

1). When the local estimates satisfy $E[x_i]=E[x]$, $i=1,2,\cdots,n$, then the fusion estimate will also satisfy $E[x^e_t]=E[x]$;

2). When the local covariances satisfy $P_i\geq E[(x_i-x)(x_i-x)^T]$, $i=1,2,\cdots,n$, then the fusion covariance will also satisfy $P^e_t\geq E[(x^e_t-x)(x^e_t-x)^T]$.

\textbf{Proof:} It is clear that the ESCI fusion (3) reduces to BCI fusion (1) when $a=\{n\}$, in which case (2) is consistent and unbiased \cite{Niehsen1020907}. Thus, only the case of $a\neq \{n\}$ will be discussed later. Firstly, it follows from (3) that
\begin{equation}\begin{aligned}
(P^e_i)^{-1}x^e_{i}=w^e_i(P^e_{i-1})^{-1}x^e_{i-1}+\sum^{b_i}_{j=b_{i-1}+1}w^{e}_{i,j}P^{-1}_{r_j}x_{r_j}
\end{aligned}\end{equation}
Then, by substituting (5) into (3), $x^e_t$ is recursively reduced to
\begin{equation}\begin{aligned}
x^e_t=&\sum^{t}_{k=2}\prod^{t}_{i=k}w^{e}_{i}P^e_t\sum^{b_{k-1}}_{j=b_{k-2}+1}w^e_{k-1,j}P^{-1}_{r_j}x_{r_j}\\ &+\sum^{b_t}_{j=b_{t-1}+1}w^{e}_{t,j}P^e_tP^{-1}_{r_j}x_{r_j}.
\end{aligned}\end{equation}
Similarly, $P^e_t$ can also be recursively simplified as
\begin{equation}\begin{aligned}
P^e_t=&(\sum^{t}_{k=2}\prod^{t}_{i=k}w^{e}_{i}\sum^{b_{k-1}}_{j=b_{k-2}+1}w^{e}_{k-1,j}P^{-1}_{r_j}\\
&+\sum^{b_t}_{j=b_{t-1}+1}w^{e}_{t,j}P^{-1}_{r_j})^{-1}.
\end{aligned}\end{equation}
Let us define
\begin{equation}\begin{aligned}
\left\{ \begin{array}{l}
u_{j}\triangleq \prod^{t}_{i=2}w^{e}_{i}w^{e}_{1,j},\ j=b_0+1,\cdots,b_1,\\
u_{j}\triangleq \prod^{t}_{i=3}w^{e}_{i}w^{e}_{2,j},\ j=b_1+1,\cdots,b_2,\\
\ \ \ \ \ \ \ \ \ \ \ \ \ \ \ \ \ \ \ \ \ \ \vdots\\
u_{j}\triangleq \prod^{t}_{i=t}w^{e}_{i}w^{e}_{t-1,j},\ j=b_{t-2}+1,\cdots,b_{t-1},\\
u_{j}\triangleq w^{e}_{t,j},\ j=b_{t-1}+1,\cdots,b_t,
\end{array} \right.
\end{aligned}\end{equation}
and then (6) and (7) are rewritten as
\begin{equation}\begin{aligned}
x^e_{t}=\sum^{n}_{i=1}u_{i}P^e_{t}P^{-1}_{r_i}x_{r_i},
\end{aligned}\end{equation}
\begin{equation}\begin{aligned}
P^e_{t}=(\sum^{n}_{i=1}u_{i}P_{r_i}^{-1})^{-1}.
\end{aligned}\end{equation}
When $E[x_i]=E[x]$, it follows from (9) and (10) that
\begin{equation}\begin{aligned}
E[x^e_{t}]=&\sum^{n}_{i=1}u_{i}P^e_{t}P^{-1}_{r_i}E[x_{r_i}]\\
=&P^e_{t}(\sum^{n}_{i=1}u_{i}P^{-1}_{r_i})E[x]=E[x].
\end{aligned}\end{equation}
Thus, the property (T.1) in Theorem 1 holds.

On the other hand, it is seen from (9) and (10) that the simplified $P^e_t$ has a similar form to $P^b$. At this point, it only needs to show that $\sum^n_{i=1}u_i=1$ and $0\leq u_i\leq 1$ for property (T.2) to hold. Therefore, substituting (4) into (8) yields
\begin{equation}\begin{aligned}
\left\{ \begin{array}{l}
\sum^{b_1}_{j=b_0+1}u_{j}=\prod^{t}_{i=2}w^{e}_{i}\sum^{b_1}_{j=b_0+1}w^{e}_{1,j}\\
\ \ \ \ \ \ \ \ \ \ \ \ \ \ \ \ \ \ =\prod^{t}_{i=2}w^{e}_{i},\\
\sum^{b_2}_{j=b_1+1}u_{j}=\prod^{t}_{i=3}w^{e}_{i}\sum^{b_2}_{j=b_1+1}w^{e}_{2,j}\\
\ \ \ \ \ \ \ \ \ \ \ \ \ \ \ \ \ \ =\prod^{t}_{i=3}w^{e}_{i}(1-w^s_2),\\
\ \ \ \ \ \ \ \ \ \ \ \ \ \ \ \ \ \ \ \ \ \ \vdots\\
\sum^{b_{t-1}}_{j=b_{t-2}+1}u_{j}=\prod^{t}_{i=t}w^{e}_{i}\sum^{b_{t-1}}_{j=b_{t-2}+1}w^{e}_{t-1,j}\\ \ \ \ \ \ \ \ \ \ \ \ \ \ \ \ \ \ \ \ \ \ =\prod^{t}_{i=t}w^{e}_{i}(1-w^e_{t-1}),\\
\sum^{b_t}_{j=b_{t-1}+1}u_{j}=\sum^{b_t}_{j=b_{t-1}+1}w^{e}_{t,j}=1-w^e_t.
\end{array} \right.
\end{aligned}\end{equation}
Then, summing over $u_i$ according to (12), one has
\begin{equation}\begin{aligned}
\sum^{n}_{j=1}u_{j}=&(1-w^e_t)+(w^e_t-w^e_tw^e_{t-1})\\
&+(w^e_tw^e_{t-1}-w^e_tw^e_{t-1}w^e_{t-2})+\cdots\\
&+(\prod_{i=3}^{t}w^e_t-\prod_{i=2}^{t}w^e_t)+\prod_{i=2}^{t}w^e_t=1.
\end{aligned}\end{equation}
Moreover, from (4) and (8) one can easily obtain $0\leq u_i\leq 1$. Under this case, it follows from the section 2 of \cite{Niehsen1020907} that the property (T.2) in Theorem 1 holds. This completes the proof. $\square$
%\subsection{Enhanced Sequential CI fusion}
%First, we make an analysis of the rationality of the proposed sequential CI fuser. Here, the rationality mainly includes unbiasedness and conservativeness, which are important properties possessed by the batch CI fuser (1).

%It is clear from Theorem 1 that the ESCI fusion framework has some similar properties to the BCI and SCI fusion frameworks, i.e., they both ensure unbiased and consistent fusion EPs. Moreover, notice that when $t=1$ and $t=n$, the proposed ESCI fusion framework (2) will degenerate to the BCI and SCI fusion frameworks, respectively. These facts indicate that the ESCI fusion framework is more universal and flexible, and therefore also has better application prospect.
Furthermore, the following theorem will give a weighting criterion for (3) such that $\{x^e_t,P^e_t\}$ is independent of the fusion structure.

%Then, the following theorem will give the weights design criteria for the ESCI fusion framework that makes $\{x^s_t,P^s_t\}$ is independent of the fusion structure.
%we will analyse the proposed sequential CI fusion framework and derive design criteria that can solve Problem 1. According to Fig. 2(b) and equation (2), it is known that the batch CI fusion is actually a special case of the proposed sequential fusion at $t=1$. Moreover, the batch CI fusion is not affected by $r_i$, $a_i$ and $t$. Therefore, the fusion result of the proposed sequential fuser (2) can always be made equivalent to the batch fuser (1) to eliminate the influence of $r_i$, $a_i$ and $t$ and to ensure the fusion performance. Then, the solution to this problem will then be given in the following theorem.
%
%\textit{Design the weights $w^s_i$ and $w^s_{i,j}$ in (2) such that the final fusion result $\{x^s_t,P^s_t\}$ is consistent with (1) regardless of how $r_i$, $a_i$ and $t$ vary. Particularly, the weight $w^b_i$ in batch fusion is considered to be known in this paper.}

\textbf{Theorem 2:} The weights $w^e_i$ and $w^e_{i,j}$ in (3) can be designed as
\begin{equation}\begin{aligned}
\left\{ \begin{array}{l}
w^e_{i}=\frac{\sum^{b_{i-1}}_{k=1}f(\{x_{r_k},P_{r_k}\})}{\sum^{b_i}_{k=1}f(\{x_{r_k},P_{r_k}\})}\\ w^e_{i,j}=\frac{f(\{x_{r_j},P_{r_j}\})}{\sum^{b_i}_{k=1}f(\{x_{r_k},P_{r_k}\})}
\end{array} \right.\,
\end{aligned}\end{equation}
such that $\{x^e_t,P^e_t\}$ is independent of $r$ and $a$. Here, $f(\{x_i,P_i\})$ is a function whose input is a EP and output is a positive scalar.

\textbf{Proof:} When $w^e_i$ and $w^e_{i,j}$ satisfy (14), it is obvious that the normalization condition in (4) holds.

By substituting (14) into (8), one has
\begin{equation}\begin{aligned}
u_i=\frac{f(\{x_{r_i},P_{r_i}\})}{\sum^{n}_{j=1}f(\{x_{r_j},P_{r_j}\})}
\end{aligned}\end{equation}
Moreover, although the fusion order $r$ is arbitrary, the available local estimates are constant, which means that $\{r_i|i=1,2,\cdots,n\}=\{i|i=1,2,\cdots,n\}$.
In this case, it follows from (15) that (9) and (10) are reduced to
\begin{equation}\begin{aligned}
x^e_{t}&=\sum^{n}_{i=1}\frac{f(\{x_{r_i},P_{r_i}\})}{\sum^{n}_{j=1}f(\{x_{r_j},P_{r_j}\})}P^e_{t}P^{-1}_{r_i}x_{r_i}\\
&=\sum^{n}_{i=1}\frac{f(\{x_i,P_i\})}{\sum^{n}_{j=1}f(\{x_{j},P_{j}\})}P^e_{t}P^{-1}_ix_i,
\end{aligned}\end{equation}
\begin{equation}\begin{aligned}
P^e_{t}&=(\sum^{n}_{i=1}\frac{f(\{x_{r_i},P_{r_i}\})}{\sum^{n}_{j=1}f(\{x_{r_j},P_{r_j}\})}P_{r_i}^{-1})^{-1}\\
&=(\sum^{n}_{i=1}\frac{f(\{x_i,P_i\})}{\sum^{n}_{j=1}f(\{x_j,P_j\})}P_i^{-1})^{-1}.
\end{aligned}\end{equation}
Obviously, the final expressions (16) and (17) do not contain $r$ and $a$, and hence $\{x^e_t,P^e_t\}$ is independent of them. The proof is completed. $\square$

%Theorem 2 gives a general criterion in which $f(\{x_{i},P_{i}\})$ does not have an explicit expression.
According to Theorem 2, $\{x^e_t,P^e_t\}$ is equal to (16) and (17) regardless of $r$ and $a$ when $w^e_i$ and $w^e_{i,j}$ ($j=b_{i-1}+1,\cdots,b_i$, $i=1,2,\cdots,t$) are calculated by (14). In this case, it is sufficient to only discuss (16) and (17) in what follows. From (16) and (17), $\{x^e_t,P^e_t\}$ is only relevant to $\{x_i,P_i\}$ and $f(\cdot)$. Therefore, it is essential to choose a reasonable $f(\cdot)$.
%Note that this paper considers the fusion estimation problem and hence $f(\cdot)$ is of primary concern, rather than the local EP $\{x_i,P_i\}$.
Then, an intuitive idea is to adopt the principle of importance assignment (i.e., assigning greater weight to $\{x_i,P_i\}$ that have a high degree of importance). Notice that, the weights in (16) and (17) satisfy
\begin{equation}\begin{aligned}
\left\{ \begin{array}{l}
f(\{x_i,P_i\})\geq f(\{x_k,P_k\})\\
\ \ \ \ \ \ \ \ \ \ \ \ \ \ \ \ \Updownarrow\\
\frac{f(\{x_i,P_i\})}{\sum^{n}_{j=1}f(\{x_j,P_j\})}\geq\frac{f(\{x_k,P_k\})}{\sum^{n}_{j=1}f(\{x_j,P_j\})}.
\end{array} \right.\,
\nonumber
\end{aligned}\end{equation}
This fact inspires us to choose $f(\{x_i,P_i\})$ as an indicator to measure the degree of importance of $\{x_i,P_i\}$. It is well known that the smaller the trace or determinant of the covariance, the more accurate the data tend to be. Therefore, it is advocated that $f(\{x_i,P_i\})$ can be chosen as $\frac{1}{\mathrm{Tr}(P_i)}$ or $\frac{1}{\mathrm{Det}(P_i)}$. Moreover, the inverse of the covariance matrix is known as the information matrix, and larger trace or determinant of the information matrix, the more reliable the data tend to be. Therefore, $f(\{x_i,P_i\})$ is also suggested to be chosen as $\mathrm{Tr}((P_i)^{-1})$ or $\mathrm{Det}((P_i)^{-1})$. Furthermore, apart from $\frac{1}{\mathrm{Tr}(P_i)}$, $\frac{1}{\mathrm{Det}(P_i)}$ and $\mathrm{Tr}((P_i)^{-1})$, $f(\{x_i,P_i\})$ can also be chosen to be other forms depending on the practical situations, as long as it reflects the degree of importance of $\{x_i,P_i\}$. For example, choose $f(\{x_i,P_i\})$ as $\frac{1}{\mathrm{Tr}(DP_i)}$, where $D$ is a diagonal matrix. When part of the components of $x$ are of primary interest, larger values can be assigned to the corresponding diagonal terms of $D$.

Based on the above analysis, we can summarize an ESCI fusion algorithm and the specific implementation steps are shown in Algorithm 1.

\begin{algorithm}
\caption{ESCI fusion algorithm}
\begin{algorithmic}[1]
\STATE  $i=0$, $t=0$, $b_0\triangleq0$, $W_0\triangleq0$, $x^e_0\triangleq O$, $(P^e_0)^{-1}\triangleq O$;
\WHILE{new local EP is received}
\STATE  $i\gets i+1$;
\STATE  Record the latest received EP as $\{x_i,P_i\}$;
\IF{Event $\mathcal{A}$ is true}
\STATE $t\gets t+1$, $b_t=i$;
%\FOR{$j=b_{t-1}+1:1:b_t$}
%\STATE $w_j=f(\{x_j,P_j\})$;
%\ENDFOR
\STATE $W_t=W_{t-1}+\sum^{b_t}_{j=b_{t-1}}w_j$, where $w_j=f(\{x_j,P_j\})$;
\STATE
$\left\{ \begin{array}{l}
x^e_t=\frac{W_{t-1}}{W_t}P^e_t(P^e_{t-1})^{-1}x^e_{t-1}\\
\ \ \ \ \ \ \ \ +\sum^{b_t}_{j=b_{t-1}}\frac{w_j}{W_t}P^e_tP^{-1}_tx_t\\ P^e_t=(\frac{W_{t-1}}{W_t}(P^e_{t-1})^{-1}+\sum^{b_t}_{j=b_{t-1}}\frac{w_j}{W_t}P^{-1}_t)^{-1}
\end{array} \right.$
\ENDIF
\ENDWHILE
\STATE Output $\{x^e_t,P^e_t\}$.
\end{algorithmic}
\label{algo:1}
\end{algorithm}

%\begin{algorithm}
%\caption{Enhanced sequential CI fusion}
%\begin{algorithmic}[1]
%\STATE  $i=0$, $t=0$, $b_0\triangleq0$, $W_0\triangleq0$, $x_{s,0}\triangleq O$, $P^{-1}_{s,0}\triangleq O$;
%\WHILE{new EPs are received}
%\STATE $t\gets t+1$;
%\STATE Determine how many EPs are received in the instant, noted as $a_t$;
%\STATE $b_t=b_{t-1}+a_t$;
%\STATE Record the latest received EPs as $\{x_i,P_i\}$, $i=b_{t-1}+1,\cdots,b_t$;
%\STATE $w_i=f(\{x_i,P_i\})$;
%\STATE $b_t=i$, $W_t=W_{t-1}+\sum^{b_t}_{j=b_{t-1}}w_j$;
%\STATE
%$\left\{ \begin{array}{l}
%x_{s,t}=\frac{W_{t-1}}{W_t}P_{s,t}P^{-1}_{s,t-1}x_{s,t-1}\\
%\ \ \ \ \ \ \ \ +\sum^{b_t}_{j=b_{t-1}}\frac{w_j}{W_t}P_{s,t}P^{-1}_tx_t\\ P_{s,t}=(\frac{W_{t-1}}{W_t}P^{-1}_{s,t-1}+\sum^{b_t}_{j=b_{t-1}}\frac{w_j}{W_t}P^{-1}_t)^{-1}
%\end{array} \right.$
%\ENDWHILE
%\STATE Output $\{x_{s,t},P_{s,t}\}$.
%\end{algorithmic}
%\label{algo:1}
%\end{algorithm}
\textbf{Remark 2:} When $\mathcal{A}$ is set to ``all local EPs have been received", the data fusion process of Algorithm 1 is the same as (1). When $\mathcal{A}$ always holds, the data fusion process of Algorithm 1 is the same as (2). Furthermore, in practice the FN may receive multiple EPs in a flash. In this case, an alternative is to set $\mathcal{A}$ as a time-related event, i.e., the FN divided the period $\Delta t$ into $m$ equal intervals, and fusion will be carried out once every $\Delta t/m$ passes, as shown in Fig. 3.
%Specifically, at $kT+iT/m$, $i=1,2,\cdots,m$, the FN will fuse the local EPs received in the time interval $(kT+(i-1)T/m,\ kT+iT/m)$ with the EP obtained from the last sequential fusion. If no EP is received in the time period, then the fusion will not be performed.
This is equivalent to set $\mathcal{A}$ as ``$t/(\Delta t/m)=0$", where $t$ represents the current time.
    \begin{figure}[thpb]
      \centering
      \includegraphics[scale=0.27]{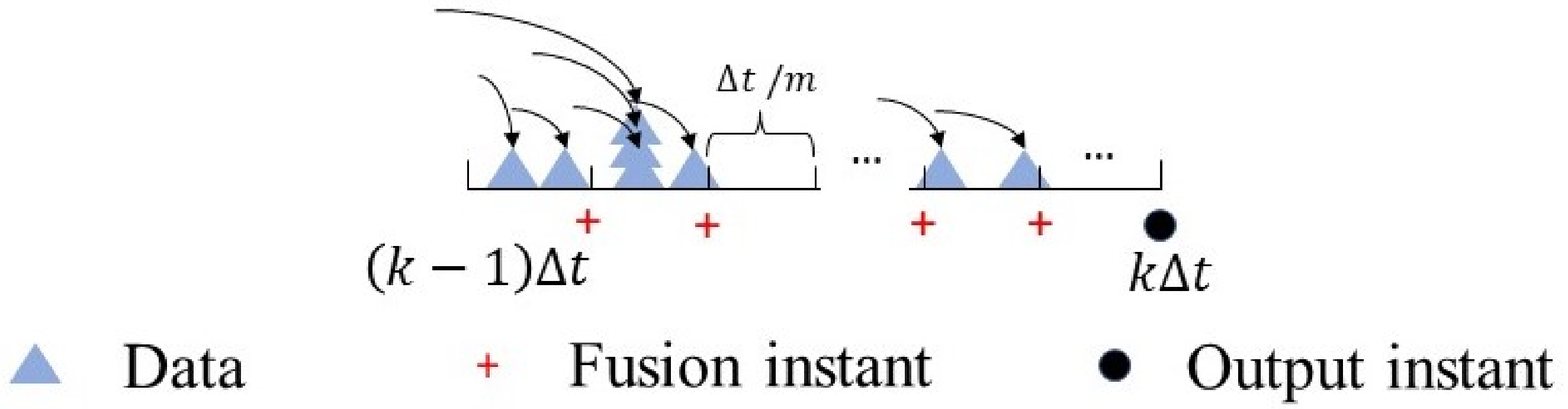}
      \caption{The time shaft corresponding to Algorithm 1 when $\mathcal{A}$ is a time-related event.}
      \label{figurelabe4}
    \end{figure}

\textbf{Remark 3:} To eliminate the effect of the fusion order $r$ on (2), a SFCI fusion algorithm was proposed in \cite{CONG201628}. Although this algorithm is independent of $r$, it measured the degree of importance of $\{x_i,P_i\}$ by $\frac{1}{\mathrm{Tr}((P_i)^{-1})}$. In this case, greater weights are assigned to those EPs with higher unreliability, which is clearly not reasonable. Notice that (2) is a special case of (3), and thus the proposed ESCI fusion algorithm can also be applied in (2) to solve the problem in \cite{CONG201628}. Furthermore, due to the more reasonable importance indicator being chosen, the ESCI fusion algorithm tends to be superior to the SFCI fusion algorithm in terms of performance, which has been shown by the simulations in Examples 2-3.

\textbf{Remark 4:} The algorithms in \cite{DENG2012293,Wang7531499,Wang8028191} do not work in the ESCI fusion framework (3) in this paper, and therefore cannot be compared with Algorithm 1. However, note that \cite{DENG2012293,Wang7531499,Wang8028191} both adopted the same scheme, i.e., the classic BCI (CBCI) fusion algorithm was directly used in the SCI fusion framework (2). Similarly, we can also use the CBCI fusion algorithm directly in (3), which is referred to as the CSCI fusion algorithm based on \cite{DENG2012293,Wang7531499,Wang8028191}. In this case, Algorithm 1 can be compared with the CSCI fusion algorithm based on \cite{DENG2012293,Wang7531499,Wang8028191} to show the advantages of the ESCI fusion algorithm.
\section{Simulation Results}
\subsection{Numerical example}
In this subsection, a intuitive numerical example is employed to show the influence of the fusion structure on different methods. Suppose that the following 4 local EPs are available:
\begin{equation}\begin{aligned}
&\{x_1,P_1\}=\{(0,-0.1),\ (2,\ 0.1;\ 0.1,\ 1.5)\}\\
&\{x_2,P_2\}=\{(-0.2,0.3),\ (3,\ 0.7;\ 0.7,\ 2)\}\\
&\{x_3,P_3\}=\{(-0.5,-0.35),\ (1.5,\ 0.5;\ 0.5,\ 3.2)\}\\
&\{x_4,P_4\}=\{(0.3,-0.15),\ (3.2,\ 2;\ 2,\ 3)\}.
\nonumber
\end{aligned}\end{equation}
Then, 10 different fusion structures are considered:

\textbf{Structure 1:} $a=\{4\}$, i.e., ESCI fusion (3) is the same as the BCI fusion (1).

\textbf{Structure 2-10:} Other fusion structures that different from structure 1.

The fusion results for the CSCI (see Remark 4) and the ESCI fusion algorithms are shown geometrically in Fig. 5 and Fig. 6, where the trajectory $X$ of the fusion ellipse is
\begin{equation}\begin{aligned}
\{X|(X-x_{fusion})^TP^{-1}_{fusion}(X-x_{fusion})=1\},
\nonumber
\end{aligned}\end{equation}
where $\{x_{fusion},P_{fusion}\}$ denotes the fusion EP. In particular, the `` $fmincon$ " function in MATLAB is utilized to solve for the weights in CSCI fusion algorithm. From Fig. 5, it can be seen that the CSCI fusion algorithm based on \cite{DENG2012293,Wang7531499,Wang8028191} is sensitive to the fusion structure. In contrast, it is known from Figs. 6 that, when different $f(\cdot)$ are chosen, the fusion ellipses of ESCI fusion algorithms do not vary with the fusion structure. This demonstrates that the proposed ESCI fusion algorithm is independent of the fusion structure, which is as expected for the proposed fusion methods.
    \begin{figure}[thpb]
      \centering
      \includegraphics[scale=0.65]{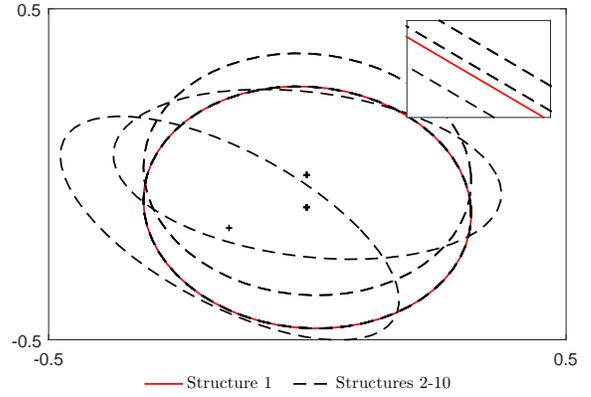}
      \caption{Fusion ellipses of CSCI fusion algotithm based on \cite{DENG2012293,Wang7531499,Wang8028191} under different fusion structures.}
      \label{figurelabe5}
    \end{figure}
    \begin{figure}[thpb]
      \centering
      \includegraphics[scale=0.6]{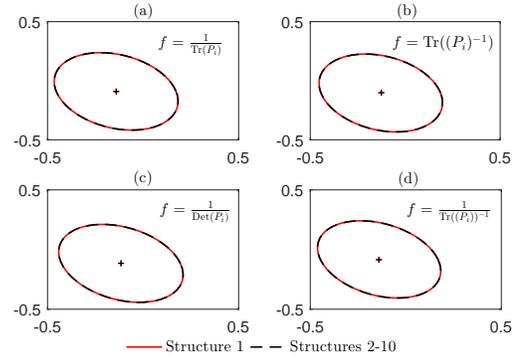}
      \caption{Fusion ellipses of ESCI fusion algotithm with different $f(\cdot)$.}
      \label{figurelabe6}
    \end{figure}
\subsection{Target Tracking System}
In this subsection, a more practical example is given to demonstrate the advantages of the ESCI fusion algorithm proposed in this paper. Consider a target tracking system with the following state-space model:
\begin{equation}\begin{aligned}
\left\{ \begin{array}{l}
x_k=\begin{bmatrix}
    {1} & {\Delta t} & {0} & {0}\\
    {0} & {\Delta t} & {0} & {0}\\
    {0} & {0} & {1} & {\Delta t}\\
    {0} & {0} & {0} & {\Delta t}
    \end{bmatrix}x_{k-1}+\begin{bmatrix}
    {\Delta t^2} & {0}\\
    {\Delta t} & {0}\\
    {0} & {\Delta t^2}\\
    {0} & {\Delta t}
    \end{bmatrix}\omega_{k-1}\\
z^i_k=\begin{bmatrix}
    {1} & {0} & {0} & {0}\\
    {0} & {0} & {1} & {0}
    \end{bmatrix}x_k+\upsilon^i_k,\ i=1,2,\cdots,10
\end{array} \right.\,
\end{aligned}\end{equation}
where $x_k=[s_{x,k}\ v_{x,k}\ s_{y,k}\ v_{y,k}]^T$. $s_{x,k}$ and $s_{y,k}$ respectively represent the coordinates of the target in the $x$ and $y$ directions. $v_{x,k}$ and $v_{y,k}$ represent the velocities of the target in the $x$ and $y$ directions, respectively. $\omega_k\in{\mathbb{R}}^2$ and $\upsilon^i_k\in{\mathbb{R}}^2$ are uncorrelated Gaussian white noise with covariance $Q_k$ and $R^i_k$, respectively. In this example, 10 sensors are used to sense the target and the Kalman filter \cite{Kalman} is embedded in each sensor to generate local EP. The simulation parameters are set as:
\begin{equation}\begin{aligned}
\left\{ \begin{array}{l}
x_0=[100m\ 10m/s\ 100m\ 5m/s],\ \Delta t=0.2s\\
Q_k=4\times\mathrm{diag}(1m^2,1m^2),\\
R^i_k=1\times\mathrm{diag}(1m^2,1m^2),\ i=1,2,3,\\
R^i_k=4\times\mathrm{diag}(1m^2,1m^2),\ i=4,5,6,\\
R^i_k=9\times\mathrm{diag}(1m^2,1m^2),\ i=7,8,9,10.
\end{array} \right.\,
\end{aligned}\end{equation}

Suppose there exists a fusion center to receive and fuse the local EPs. Meanwhile, the fusion center divides each period $\Delta t$ into 10 equal intervals (see Remark 2). Moreover, the `` $rand$ " function in MATLAB is utilized to model the unpredictability of the moments in which fusion center receives the local EPs. Then, by implementing Algorithm 1 with different $f(\cdot)$, the estimated target trajectories and actual target trajectory are drawn in Fig. 7. From this figure, it is observed that the proposed ESCI fusion algorithm can track the target well.
    \begin{figure}[thpb]
      \centering
      \includegraphics[scale=0.6]{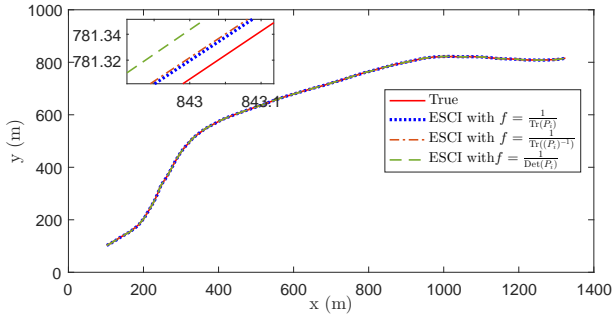}
      \caption{True trajectory and estimated trajectories by using ESCI fusion algorithms with different $f(\cdot)$.}
      \label{figurelabe7}
    \end{figure}

To compare the ESCI fusion algorithm proposed in this paper with the methods in [20-23], root mean square error (RMSE) is used as an indicator and $500$ Monte Carlo simulations are performed to approximate the ideal RMSE. Meanwhile, the CBCI fusion algorithm in \cite{Julier609105} which can minimize the trace of $P^b$ in (1) is also considered here as a benchmark. Then, Fig. 8 shows the position RMSEs of CBCI, CSCI and ESCI fusion algorithms with different $f(\cdot)$. From Fig. 8 we see that the ESCI fusion algorithm with $f=\frac{1}{\mathrm{Tr}((P_i)^{-1})}$ has the lowest accuracy, which implies that treating $\frac{1}{\mathrm{Tr}((P_i)^{-1})}$ as a importance indicator as in \cite{CONG201628} leads to poor fusion performance. This is in line with the analysis in Remark 3. Moreover, it can be known from Fig. 8 that the accuracy of CSCI fusion algorithm based on \cite{DENG2012293,Wang7531499,Wang8028191} is lower than that of CBCI fusion algorithm. This illustrates that using the CBCI fusion algorithm directly in the ESCI fusion framework is an unsatisfactory scheme (see Remark 4). Meanwhile, when $f=\frac{1}{\mathrm{Tr}(P_i)}$, $f=\frac{1}{\mathrm{Det}(P_i)}$ and $f=\mathrm{Tr}((P_i)^{-1})$, though the accuracy of ESCI fusion algorithms is slightly lower than that of CBCI fusion algorithm in \cite{Julier609105}, it is much higher than that of CSCI fusion algorithm based on \cite{DENG2012293,Wang7531499,Wang8028191}.
%Notice that, the CBCI fusion algorithm requires an optimization algorithm to solve for the weights at each moment, and it adopts the batch processing scheme, which makes it computationally inefficient. In contrary, the ESCI fusion algorithm has an explicit form for the weights which is computationally efficient. In this case, the ESCI fusion algorithm is considered to be superior to CSCI and CBCI fusion algorithms.
    \begin{figure}[thpb]
      \centering
      \includegraphics[scale=0.6]{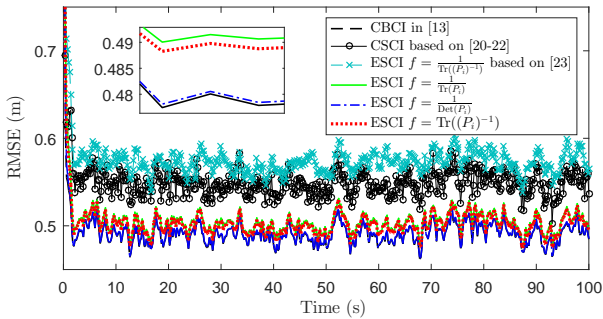}
      \caption{Position RMSEs of CBCI in \cite{Julier609105}, CSCI based on \cite{DENG2012293,Wang7531499,Wang8028191} and the proposed ESCI with different $f(\cdot)$.}
      \label{figurelabe8}
    \end{figure}

On the other hand, although CBCI fusion algorithm has the highest accuracy, its computational costs are very large and concentrated in a instant within each period, as shown in Fig. 9. Under this case, the CBCI fusion algorithm in \cite{Julier609105} tends to induce severe computational delay. Moreover, although the computational costs of CSCI fusion algorithm based on \cite{DENG2012293,Wang7531499,Wang8028191} spread over the entire period, it also requires an optimization algorithm to solve for the weights, which makes it computationally intensive as well. Furthermore, it is clear from Fig. 9 that the computational costs of the proposed ESCI fusion algorithm are small and spread over the entire time shaft, and thus the proposed ESCI fusion algorithm is more efficient.
%Therefore, when the appropriate indicator $f(\cdot)$ is chosen, ESCI fusion algorithm has high computational efficiency while guarantees the fusion performance is very close to that of CBCI fusion algorithm in \cite{Julier609105}. In this case, the ESCI fusion algorithm
    \begin{figure}[thpb]
      \centering
      \includegraphics[scale=0.6]{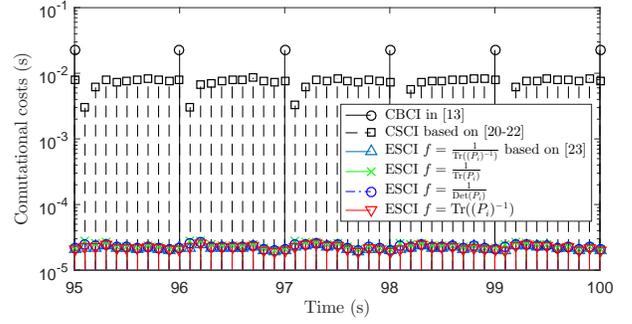}
      \caption{The distribution of computational costs for CBCI, CSCI and ESCI fusion algorithms.}
      \label{figurelabe9}
    \end{figure}
\subsection{Mobile Robot Localization}
In this subsection, the performance of the ESCI fusion algorithm in nonlinear system is further demonstrated by a common two-wheel differential robot motion model \cite{Chen8391738}:
\begin{equation}\begin{aligned}
x_{k+1}=
   \left[\begin{matrix}
   s_{x,k}+u_{v,k}\mathrm{cos}(\theta_k)\Delta t\\
   s_{y,k}+u_{v,k}\mathrm{sin}(\theta_k)\Delta t\\
   \theta_{k}+u_{\theta,k}\Delta t
   \end{matrix}\right]+\omega_k,\nonumber
\end{aligned}
\end{equation}
where the state $x_k=[s_{x,k},s_{y,k},\theta_k]$. $\theta_k$ is the heading angle, $u_{v,k}$ and $u_{\theta,k}$ are the control inputs, $w_{k}$ is Gaussian white noise with covariance $Q_k$. Four sensors equipped with the cubature Kalman filter (CKF) \cite{CKF} are used to locate the robot, and the measurement equations are
\begin{equation}\begin{aligned}
z^i_{k+1}=\left[\begin{matrix}
   \sqrt{(s_{x,k+1}-l^i_x)^2+(s_{y,k+1}-l^i_y)^2}\ \\
   \mathrm{arctan}(\frac{s_{y,k+1}-l^i_y}{s_{x,k+1}-l^i_x})
   \end{matrix}\right]+\upsilon^i_{k+1},\nonumber
\end{aligned}\end{equation}
where $(l^i_x,l^i_y)$ represents the coordinate of the $i$th sensor in the $x$-$y$ plane. $\upsilon^i_{k+1}$ is measurement noise with covariance $R^i_k$. The simulation parameters are set as:
\begin{equation}\begin{aligned}
\left\{ \begin{array}{l}
x_0=[200cm\ 200cm\ 0],\ \Delta t=0.08s,\\
Q_k=\Delta t^2\times\mathrm{diag}(1cm^2/s^2,1cm^2/s^2,(1^{\circ})^2/s^2),\\
R^i_k=0.1^2\times\mathrm{diag}(1cm^2,(1^{\circ})^2),\ i=1,2,\\
R^i_k=0.2^2\times\mathrm{diag}(1cm^2,(1^{\circ})^2),\ i=3,4.
\end{array} \right.\,
\end{aligned}\end{equation}

By implementing Algorithm 1 with different $f(\cdot)$, the estimated target trajectories and actual target trajectory are drawn in Fig. 10. From this figure, we can see that the proposed ESCI fusion algorithm provides an accurate location for the robot.
    \begin{figure}[thpb]
      \centering
      \includegraphics[scale=0.6]{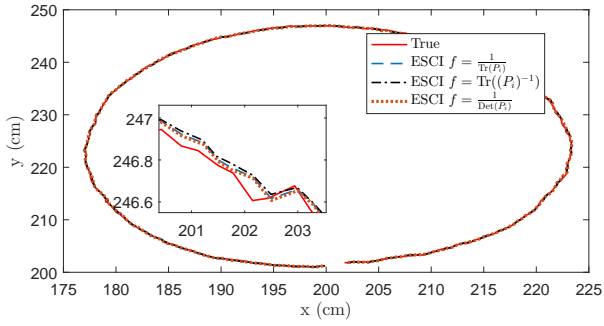}
      \caption{True trajectory and estimated trajectories by using ESCI fusion algorithms with $f=1/\mathrm{Tr}(P_i)$, $1/\mathrm{Det}(P_i)$, $\mathrm{Tr}((P_i)^{-1})$.}
      \label{figurelabe10}
    \end{figure}

Under 20 different fusion structures, the position RMSEs of the CSCI and ESCI fusion algorithms with different $f(\cdot)$ are plotted in Fig. 11. As can be seen from Fig. 11, the fusion structure has strong influences on the CSCI fusion algorithm based on \cite{DENG2012293,Wang7531499,Wang8028191}. In this case, the performance of CSCI fusion algorithm cannot be guaranteed in practice. On the contrary, it is seen from Fig. 11 that the position RMSEs of the ESCI fusion algorithms are invariant under different fusion structures. This indicates that the ESCI fusion algorithm is also independent of the fusion structure in the nonlinear system. Furthermore, it can be known from Fig. 11 that the accuracy of ESCI with $f=\frac{1}{\mathrm{Tr}(P_i)^{-1}}$ \cite{CONG201628} is lower than that of ESCI with $f=\frac{1}{\mathrm{Tr}(P_i)}$, $f=\frac{1}{\mathrm{Det}(P_i)}$, $f=\frac{1}{\mathrm{Tr}((P_i)^{-1})}$. This is because the unreliable local EPs will be assigned greater weights when the degree of importance of local EPs is measured by $\frac{1}{\mathrm{Tr}(P_i)^{-1}}$ as in \cite{CONG201628}.
    \begin{figure}[thpb]
      \centering
      \includegraphics[scale=0.6]{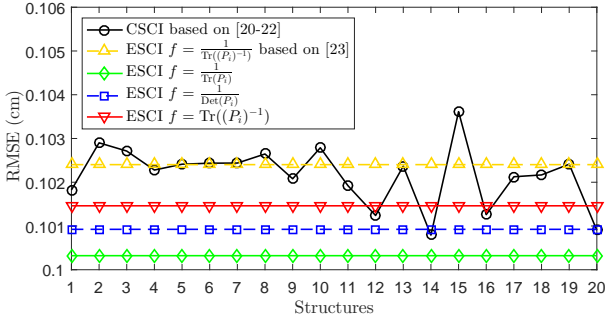}
      \caption{Position RMSEs of CSCI and the proposed ESCI fusion algorithm with different $f(\cdot)$ under different fusion structures.}
      \label{figurelabe11}
    \end{figure}

\section{Conclusion}
This paper proposed an ESCI fusion which possessed the unbiasedness and consistency. Particularly, both the existing BCI and SCI fusion were special cases of the ESCI fusion and therefore it had better prospects in practice. Notice that the fusion structure of ESCI fusion is unpredictable due to practical factors. To avoid unpredictable effects of the fusion structure on performance, a weighting criterion for the proposed ESCI fusion was further designed, which had two important properties: 1). It made the fusion results be independent of the fusion structure and thus can solve the above problem; 2). It obeyed the principle of importance assignment, and the importance indicator was not fixed but allowed to be chosen as different forms for meeting different practical requirements. Finally, three simulation examples were used to demonstrate the effectiveness and advantages of the proposed approach.
% or

  % for no appendix heading


%\appendices

%\section{}

%\section*{Acknowledgment}


\ifCLASSOPTIONcaptionsoff
  \newpage
\fi

\normalem
\bibliographystyle{ieeetr}
%\bibliography{refs}

\begin{thebibliography}{10}

\bibitem{9057730}
T.~Li, X.~Wang, Y.~Liang, and Q.~Pan, ``On arithmetic average fusion and its
  application for distributed multi-bernoulli multitarget tracking,'' {\em IEEE
  Transactions on Signal Processing}, vol.~68, pp.~2883--2896, 2020.

\bibitem{DFKLF}
Y.~Zhang, B.~Chen, and L.~Yu, ``Distributed fusion kalman filtering under
  binary sensors,'' {\em International Journal of Robust and Nonlinear
  Control}, vol.~30, no.~6, pp.~2570--2578, 2020.

\bibitem{LIN2017560}
C.~Lin, H.~Mu, R.~Xiong, and J.~Cao, ``Multi-model probabilities based state
  fusion estimation method of lithium-ion battery for electric vehicles:
  State-of-energy,'' {\em Applied Energy}, vol.~194, pp.~560--568, 2017.

\bibitem{ZHANG2018358}
W.-A. Zhang and L.~Shi, ``Sequential fusion estimation for clustered sensor
  networks,'' {\em Automatica}, vol.~89, pp.~358--363, 2018.

\bibitem{SUN20041017}
S.-L. Sun and Z.-L. Deng, ``Multi-sensor optimal information fusion kalman
  filter,'' {\em Automatica}, vol.~40, no.~6, pp.~1017--1023, 2004.

\bibitem{LIN2019128}
H.~Lin and S.~Sun, ``Globally optimal sequential and distributed fusion state
  estimation for multi-sensor systems with cross-correlated noises,'' {\em
  Automatica}, vol.~101, pp.~128--137, 2019.

\bibitem{YAN20133607}
L.~Yan, X.~{Rong Li}, Y.~Xia, and M.~Fu, ``Optimal sequential and distributed
  fusion for state estimation in cross-correlated noise,'' {\em Automatica},
  vol.~49, no.~12, pp.~3607--3612, 2013.

\bibitem{Zhang8776655}
W.-A. Zhang, K.~Zhou, X.~Yang, and A.~Liu, ``Sequential fusion estimation for
  networked multisensor nonlinear systems,'' {\em IEEE Transactions on
  Industrial Electronics}, vol.~67, no.~6, pp.~4991--4999, 2020.

\bibitem{Reinhardt7005422}
M.~Reinhardt, B.~Noack, P.~O. Arambel, and U.~D. Hanebeck, ``Minimum covariance
  bounds for the fusion under unknown correlations,'' {\em IEEE Signal
  Processing Letters}, vol.~22, no.~9, pp.~1210--1214, 2015.

\bibitem{Niehsen1020907}
W.~Niehsen, ``Information fusion based on fast covariance intersection
  filtering,'' in {\em Proceedings of the Fifth International Conference on
  Information Fusion. FUSION 2002. (IEEE Cat.No.02EX5997)}, vol.~2,
  pp.~901--904 vol.2, 2002.

\bibitem{Chen1047015}
L.~Chen, P.~Arambel, and R.~Mehra, ``Estimation under unknown correlation:
  covariance intersection revisited,'' {\em IEEE Transactions on Automatic
  Control}, vol.~47, no.~11, pp.~1879--1882, 2002.

\bibitem{Marc7005422}
M.~Reinhardt, B.~Noack, P.~O. Arambel, and U.~D. Hanebeck, ``Minimum covariance
  bounds for the fusion under unknown correlations,'' {\em IEEE Signal
  Processing Letters}, vol.~22, no.~9, pp.~1210--1214, 2015.

\bibitem{Julier609105}
S.~Julier and J.~Uhlmann, ``A non-divergent estimation algorithm in the
  presence of unknown correlations,'' in {\em Proceedings of the 1997 American
  Control Conference (Cat. No.97CH36041)}, vol.~4, pp.~2369--2373 vol.4, 1997.

\bibitem{Franken1591849}
D.~Franken and A.~Hupper, ``Improved fast covariance intersection for
  distributed data fusion,'' in {\em 2005 7th International Conference on
  Information Fusion}, vol.~1, pp.~7 pp.--, 2005.

\bibitem{Hurley1021196}
M.~Hurley, ``An information theoretic justification for covariance intersection
  and its generalization,'' in {\em Proceedings of the Fifth International
  Conference on Information Fusion. FUSION 2002. (IEEE Cat.No.02EX5997)},
  vol.~1, pp.~505--511 vol.1, 2002.

\bibitem{Wang6129634}
Y.~Wang and X.~R. Li, ``Distributed estimation fusion with unavailable
  cross-correlation,'' {\em IEEE Transactions on Aerospace and Electronic
  Systems}, vol.~48, no.~1, pp.~259--278, 2012.

\bibitem{BATTISTELLI2014707}
G.~Battistelli and L.~Chisci, ``Kullback–leibler average, consensus on
  probability densities, and distributed state estimation with guaranteed
  stability,'' {\em Automatica}, vol.~50, no.~3, pp.~707--718, 2014.

\bibitem{Duan6916177}
Z.~Duan, X.~R. Li, and U.~D. Hanebeck, ``Multi-sensor distributed estimation
  fusion using minimum distance sum,'' in {\em 17th International Conference on
  Information Fusion (FUSION)}, pp.~1--8, 2014.

\bibitem{Tang8519318}
M.~Tang, Y.~Rong, J.~Zhou, and X.~R. Li, ``Information geometric approach to
  multisensor estimation fusion,'' {\em IEEE Transactions on Signal
  Processing}, vol.~67, no.~2, pp.~279--292, 2019.

\bibitem{DENG2012293}
Z.~Deng, P.~Zhang, W.~Qi, J.~Liu, and Y.~Gao, ``Sequential covariance
  intersection fusion kalman filter,'' {\em Information Sciences}, vol.~189,
  pp.~293--309, 2012.

\bibitem{Wang7531499}
J.~Wang, Y.~Gao, and C.~Ran, ``Sequential covariance intersection fusion kalman
  filter for multi-sensor systems with multiple time delayed measurements,'' in
  {\em 2016 Chinese Control and Decision Conference (CCDC)}, pp.~3018--3022,
  2016.

\bibitem{Wang8028191}
J.~Wang, T.~Shang, Y.~Gao, C.~Ran, Y.~Huo, G.~Hao, and Y.~Li, ``Sequential
  covariance intersection fusion kalman filter for multiple time-delay sensor
  network systems with colored noises,'' in {\em 2017 36th Chinese Control
  Conference (CCC)}, pp.~5282--5287, 2017.

\bibitem{CONG201628}
J.~Cong, Y.~Li, G.~Qi, and A.~Sheng, ``An order insensitive sequential fast
  covariance intersection fusion algorithm,'' {\em Information Sciences},
  vol.~367-368, pp.~28--40, 2016.

\bibitem{Kalman}
R.~E. Kalman, ``A new approach to linear filtering and prediction problems,''
  {\em Journal of Basic Engineering}, vol.~82D, pp.~35--45, 1960.

\bibitem{Chen8391738}
B.~Chen, G.~Hu, D.~W. Ho, and L.~Yu, ``A new approach to linear/nonlinear
  distributed fusion estimation problem,'' {\em IEEE Transactions on Automatic
  Control}, vol.~64, no.~3, pp.~1301--1308, 2019.

\bibitem{CKF}
I.~Arasaratnam and S.~Haykin, ``Cubature kalman filters,'' {\em IEEE
  Transactions on Automatic Control}, vol.~54, no.~6, pp.~1254--1269, 2009.

\end{thebibliography}

\end{document}